\documentstyle[draft]{mn}

\title[Maser emission towards R~Aqr and H1$-$36]
{The simultaneous detection of 22- and 321-GHz H$_{2}$O maser emission
towards the symbiotic Mira R~Aquarii}

\author[R.\,J.\ Ivison et al.]
       {R.\,J.\ Ivison$^1$, J.\,A.\ Yates$^2$ and P.\,J.\ Hall$^3$\\
        $^1$ Institute for Astronomy, University of Edinburgh, Blackford Hill,
             Edinburgh EH9 3HJ\\
        $^2$ Division of Physics and Astronony, Dept of Physical Sciences,
             University of Hertfordshire, College Lane, Hatfield AL10 9AB\\
        $^3$ Australia National Telescope Facility, CSIRO, PO Box 76, Epping,
             NSW 2121, Australia}

\date{Accepted ...
      Received ...
      in original form 1997 June 24}
\pagerange{000--000}

\begin{document}

\maketitle

\begin{abstract}
We report high spatial and spectral resolution measurements of masers
towards R~Aqr and H1$-$36, both of which are examples of the sub-class
of symbiotic stars that contain a long-period Mira-type variable. Our
observations have resulted in the first detection of 321-GHz H$_{2}$O
maser action towards a symbiotic Mira --- R~Aqr. Comparison with
simultaneous 22-GHz H$_{2}$O maser data suggests the masers do not
have the same properties as those in the circumstellar envelopes of
field Miras. R~Aqr's 22-/321-GHz peak flux density and luminosity
ratios are low, as is the line-width ratio. Continuum and
spectral-line maps indicate that the 22-GHz maser and free-free
emission are aligned. Three mechanisms can reproduce the data with
varying degrees of success.  All three lead naturally to normal levels
of maser emission in SiO and 321-GHz H$_2$O and anomalously weak OH
and 22-GHz H$_2$O masers. In the most convincing model, UV radiation
and a fast wind from the companion remove the Mira's envelope of
dusty, molecular gas, leaving a relatively small cavity of dense,
neutral material within a large, ionized nebula.  Excitation
temperatures suggest that 321-GHz masers are normally excited close to
the Mira whilst 22-GHz masers are more remote; in R~Aqr, therefore,
the 22-GHz masers do not form under optimum conditions. Instead, we
see weak and narrow lines that form closer to the Mira, consistent
with our high-resolution maps.
\end{abstract}


\section{Introduction}

R~Aqr is the closest known symbiotic Mira. It is a binary system in
which a late-type, long-period variable giant loses matter to a
post-AGB dwarf.  Existing models have failed to account convincingly
for many of the simplest features of the R~Aqr system: the binary
period, the method of mass transfer, the nature of the hot, companion
star and its outburst mechanism. Still less is known about the more
unusual phenomenon on display: the transient jet-like prominence
(Wallerstein \& Greenstein 1980; Sopka et al.\ 1982) and the
arcmin-scale nebulae (Solf \& Ulrich 1985).

Various studies have shown that the molecular emission lines of
symbiotic stars are very different from those expected from isolated
Miras. For example, Bowers \& Hagen (1984) stated that 75 per cent of
Miras in the solar neighbourhood have 22-GHz H$_2$O maser emission,
whereas Seaquist, Ivison \& Hall (1995 --- SIH95) reported that only
two of the twenty-five known Mira-type symbiotic stars support 22-GHz
H$_2$O masers. R~Aqr and H1$-$36 are the unusual duo. The reasons why
these systems are peculiar are not well understood, but it seems
likely that it is because of the relatively large size of the binary
orbit in the case of H1$-$36, and because of the low luminosity of the
hot companion star in the case of R~Aqr (compared with other symbiotic
objects in the samples observed by SIH95 and Schwarz et al.\ 1995).

 \begin{table*}
 \caption{Parameters for four epochs of 22-GHz observations of R~Aqr.}
 \label{raqr-vla}
 \begin{tabular}{lccccc}
 \hline
UT Date    &Array        &Synthesized beam&Dominant peak          &\multicolumn{2}{c}{Position (B1950)}\\
           &configuration&size and PA     &velocity (km\,s$^{-1}$)&$\alpha (23^{\rm h} 41')$&$\delta (-15^{\circ} 33')$\\
&&&&&\\
1993 May 16&CnB          &$1.8'' \times 0.6'', 124^{\circ}$&$-26.6\pm0.2$&$14.27\pm0^{\rm s}.02$  &$43.30\pm0''.15$\\
1994 May 20&BnA          &$0.2'' \times 0.1'', 55^{\circ}$ &$-26.8\pm0.2$&$14.27\pm0^{\rm s}.01$  &$43.35\pm0''.03$\\
1995 June 8&D            &$4.3'' \times 2.6'', 154^{\circ}$&$-24.7\pm0.2$&$14.26\pm0^{\rm s}.03$  &$43.19\pm0''.60$\\
 \hline
 \end{tabular}
 \end{table*}

R~Aqr and H1$-$36 also harbour amongst the brightest known SiO
masers. R~Aqr has been detected in the 43-GHz $J=1-0$ lines of both
the $v=1$ and $v=2$ states (Martinez et al.\ 1988), with SiO fluxes at
43\,GHz averaging around 100\,Jy over several pulsational periods,
with the flux and velocity of individual features varying considerably
from one period to the next. It has also been detected in the $v=1$,
$J=2-1$ transition at 86\,GHz (Hollis et al.\ 1990). Schwarz et al.\
show that the peak of the 86-GHz emission averages about 200\,Jy, but
varies with stellar phase and between different periods in both flux
and velocity (like the 43-GHz emission). Finally, R~Aqr is also one of
a handful of systems to display the $J=7-6$ SiO transition (Gray et
al.\ 1995).
 
In a recent survey of twenty-two circumstellar envelope 22-GHz H$_2$O
maser sources, Yates, Cohen and Hills (1995) showed that 70 per cent
had observable sub-mm H$_2$O emission.  The velocity extent of 321-GHz
maser emission was less than that at 22-GHz and the 22-GHz peak flux
density was typically twice that at 321-GHz.

Here, we report the first detection of the $J=10_{29} - 9_{36}$ ortho
321-GHz H$_2$O maser transition towards a symbiotic Mira system. Poor
observing conditions prevented any attempt to detect the para
$J=5_{15} - 4_{22}$ 325-GHz H$_2$O maser transition which sits in a
deep atmospheric absorption trough.  Our sub-mm observations were
performed using the James Clerk Maxwell Telescope, whilst observations
of the $J=6_{16} - 5_{23}$ 22-GHz H$_2$O maser transition were
obtained contemporaneously using the Very Large Array.

\section{Observations}

\subsection{James Clerk Maxwell Telescope}

The JCMT observations were carried out at two epochs. A
liquid-helium-cooled, single-channel SIS mixer receiver, B3i, was
used, with a digital autocorrelation spectrometer backend providing a
bandwidth of 250\,MHz and a channel spacing of 0.156\,MHz. The rest
frequency adopted for the H$_2$O $J=10_{29} - 9_{36}$ transition was
321.22564\,GHz, with velocities of $-25.0$ and $-125.0$\,km\,s$^{\rm
-1}$ relative to the LSR for R~Aqr and H1$-$36, respectively.

During 1994 May 01, observations were hampered by weather conditions
unsuited to observing near the 325-GHz atmospheric water absorption
feature --- the opacity at 230\,GHz was $\sim 0.1$, giving a 321-GHz
system temperature of around 1500\,{\sc k}. We integrated for 1\,hr,
achieving a 3-$\sigma$ upper limit on the integrated line flux of
13.8\,Jy\,km\,s$^{-1}$.
 
At the second epoch, 1995 June 05, $T_{\rm sys}$ was $\sim1000$\,{\sc
k}.  After integrating for 5.25\,hr (exclusive of overheads such as
the time spent nodding the telescope to alternate the signal and
reference beams), we obtained a detection of the 321-GHz transition
towards R~Aqr.

\subsection{Very Large Array}

\begin{figure}
\setlength{\unitlength}{1mm}
\begin{picture}(50,117)
\put(0,0){\includegraphics{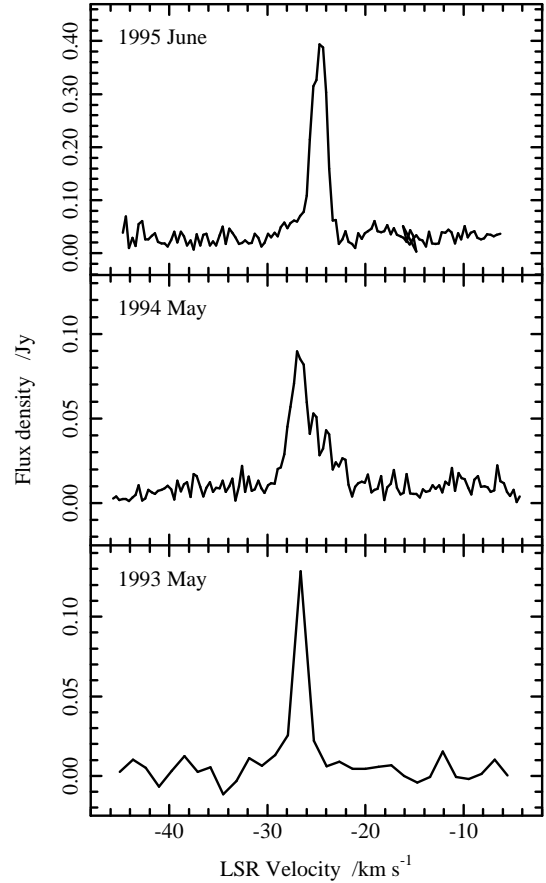}}
\end{picture}
\caption{The 22-GHz spectra observed towards R~Aqr using the VLA
during 1993 May (lower, from Ivison et al.\ 1994), 1994 May (middle)
and 1995 June (upper). Pulsational phases, where $\phi=0.00$
corresponds to maximum visual light, are $\phi = 0.33$, $0.28$ and $0.27$
(Chinarova, Andronov \& Schweitzer 1996). Improved velocity resolution
is apparent in the upper panels.}
\label{vla}
\end{figure}

\begin{figure}
\setlength{\unitlength}{1mm}
\begin{picture}(200,72)
\put(0,0){\includegraphics{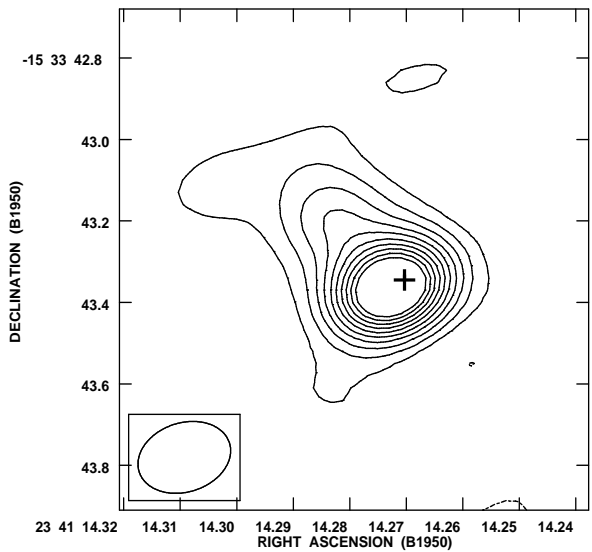}}
\end{picture}
\caption{BnA-configuration continuum map of R~Aqr at 22\,GHz. The map
is marked with the position of the 22-GHz H$_2$O maser and a
representation of the restoring beam. The peak flux density is
12.5\,mJy\,beam$^{-1}$. Contours are plotted at integer values of the
3-$\sigma$ level (0.85\,mJy\,beam$^{-1}$). See Ivison et al.\ (1995a)
for a line-free map extracted from the 1665- and 1667-MHz OH
data cubes with the position of the 22-GHz H$_2$O maser marked.}
\label{maps}
\end{figure}

Including data reported by Ivison, Seaquist \& Hall (1994),
interferometric observations at 22\,GHz have been carried out using
the VLA on four occasions. Here, we describe the latter three epochs.

The telescope was in the A configuration during 1994 May 01-02 when
data were obtained for H1$-$36 and R~Aqr covering the 1612- and
1667-MHz OH lines simultaneously with two IF pairs (for H1$-$36:
195-kHz bandwidth, with 0.28-km\,s$^{-1}$ channel separation, for
150\,min; for R~Aqr: 781-kHz bandwidth, with 1.13-km\,s$^{-1}$ channel
separation, for 120\,min). Data was also obtained for H1$-$36 covering
the 22-GHz H$_2$O line (6.25-MHz bandwidth, with 1.32-km\,s$^{-1}$
channel separation, for 45\,min). The H1$-$36 data were phase
referenced against 1748$-$253 and 1741$-$312 at the low and high
frequencies, respectively, whilst the R~Aqr data were referenced
against 2345$-$167.

During 1994 May 20, the VLA was used in BnA configuration (i.e.\ B
configuration, with some long N-S baselines). Data were obtained for
R~Aqr covering the 22-GHz H$_2$O line (3.125-MHz bandwidth, with
0.33-km\,s$^{-1}$ channel separation, for 140\,min). A 22-GHz
continuum map of R~Aqr, with 100-MHz bandwidth, was also obtained.

Our final set of VLA observations were performed, using the D
configuration, during 1995 June 08 and were contemporaneous with the
JCMT measurements described earlier. Data were obtained covering the
22-GHz H$_2$O line for both H1$-$36 and R~Aqr (1\,hr each, after
overheads) with the same bandwidths and channel separations as used in
1994 May.

For all the VLA observations, absolute flux calibration was performed
using 0134$+$329 and 1328$+$307.

\section{Observations of R~Aquarii}

\subsection{The H$_2$O masers}

\begin{figure}
\setlength{\unitlength}{1mm}
\begin{picture}(50,79)
\put(0,0){\includegraphics{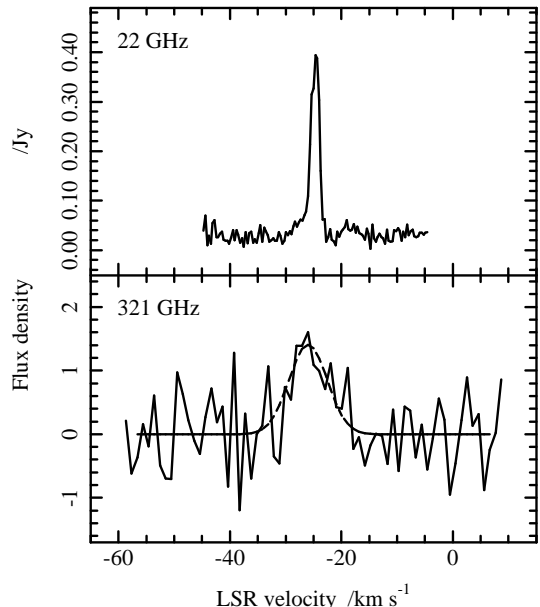}}
\end{picture}
\caption{The 22-GHz (upper) and 321-GHz (lower) spectra observed towards
R~Aqr during 1995 June ($\phi = 0.27$). The 321-GHz data from JCMT
have been fitted with a Gaussian profile and are shown after
subtraction of a 2nd-order polynomial fit to the baseline.}
\label{1995}
\end{figure}

Fig.~\ref{vla} shows all the observations of R~Aqr obtained so far
with the VLA at 22\,GHz. During our initial maser-line survey
(reported by Ivison et al.\ 1994; SIH95), a relatively poor velocity
resolution was employed, both to improve our sensitivity to weak lines
and to maximise velocity coverage. At subsequent epochs we used
narrower channels, which is apparent from the improved velocity
resolution of spectra in the upper panels of Fig.~\ref{vla}.

The 22-GHz H$_2$O line from R~Aqr shows temporal variations similar to
those seen in its SiO maser lines (Gray et al.\ 1998).  The full
velocity width of the 22-GHz H$_2$O line (8\,km\,s$^{-1}$) is also the
same as that of the 86-GHz $v=1, J=2-1$ SiO line observed by Schwarz
et al.\ (1995), confirming the speculation of Ivison et al.\ (1994).

In terms of its position, the H$_2$O maser is coincident with the
centroid of the 22-GHz continuum emission (Fig.~\ref{maps})
($<50$\,milliarcseconds, or $< 11$\,au for a distance of 220\,pc ---
van Belle et al.\ 1996).  These data were obtained contemporaneously.
During 1994 May, the full velocity extent of 8\,km~s$^{-1}$ and the
presence of several spectral features suggests that the 22-GHz maser
region of R~Aqr is complex and made up of several discrete masing
clumps. The 0.22 $\times$ 0.11 mas$^{2}$ restoring beam did not
resolve any structure, but it does allow us to confirm that the
emission mechanism is definitely non-thermal, since $T_{\rm b}$ $>$
12,000\,{\sc k}.

At the other two epochs, there was one dominant peak. The velocities
of the dominant line peaks during 1993 and 1994 were separated by only
0.2\,km\,s$^{-1}$, but a 2.1-km\,s$^{-1}$ redward shift is evident in
the 1995 data. The full profile width is similar at all three epochs,
always covering the $-22$ to $-30$\,km\,s$^{-1}$ $v_{\rm lsr}$
interval.

Fig.~\ref{1995} shows the 22- and 321-GHz spectra obtained during 1995
June for R~Aqr. The 321-GHz spectrum is noisy, so it would be a
mistake to draw too many conclusions from the line profile or
width. The best Gaussian fit to the data yields a FWHM line width of
$8\pm2$\,km\,s$^{-1}$ with a central velocity of
$-26\pm1$\,km\,s$^{-1}$ (slightly blueward and broader than the
dominant 22-GHz line peak --- see below). The integrated flux density
is $10.5\pm2.5$\,Jy\,km\,s$^{-1}$ and the corresponding photon
luminosity is $(3.06 \pm 0.73) \times 10^{41}$\,s$^{-1}$.

The upper plot of Fig.~\ref{1995} shows the contemporanous 22-GHz
spectrum. As with all our VLA data, the line emission stretches from
$-22$ to $-30$\,km\,s$^{-1}$; the dominant peak (0.37\,Jy after
subtracting off the free-free continuum contribution) is at $v_{\rm
lsr} = -24.7$\,km\,s$^{-1}$; its FWHM is around
$1.8\pm0.1$\,km\,s$^{-1}$ whilst the integrated flux density of the
entire profile, including the pedestal, is
$0.73\pm0.05$\,Jy\,km\,s$^{-1}$ or $(6.4 \pm 0.4) \times
10^{40}$\,photon\,s$^{-1}$.

\subsection{Comparision with the H$_2$O masers of field Miras}

The recent survey of the profile characteristics of 22- and 321-GHz
maser emission from twenty field Miras (Yates et al.\ 1995) gave a
mean ratio of $2.2$ for the peak flux densities of the 22- and 321-GHz
lines and a mean luminosity ratio of $1.5$. There was no apparent
correlation between these ratios and the mass-loss rate or the {\em
IRAS} colours, though the line luminosities did correlate with both of
these parameters.

For R~Aqr we measured a peak flux density ratio of $0.24\pm0.05$ and a
photon luminosity ratio of $0.21\pm0.05$. These lie at the lowest
bound of the values observed by Yates et al.\ (1995). The only objects
in the Yates et al.\ sample that rival these ratios are o~Ceti, U~Her,
R~Leo and R~Cas; the former is mildly symbiotic -- a wind-accreting
companion star is separated from the Mira by around 80\,au (Karovska,
Nisenson \& Belefic 1993); all four have relatively low mass-loss
rates. R~Aqr also has a very low mass-loss rate ($0.06 \times
10^{-6}$\,M$_{\odot}$\,yr$^{-1}$, Seaquist, Krogulec \& Taylor 1993).

As was mentioned earlier, Yates et al.\ (1995) found no correlation
between the 22-/321-GHz maser-line photon luminosity ratio and the
mass-loss rate for a sample of 13.  We have re-analysed the data
presented by Yates et al., supplementing them with the result for
R~Aqr and treating upper and lower limits for the 22-/321-GHz
maser-line photon luminosity ratio as though they were detections (the
effect of which is to underestimate the degree of correlation, since
the systems with low mass-loss rates are those with upper limits and
the systems with high mass-loss rates are those with lower
limits). With a sample of 19, the resulting correlation coefficient is
0.66, indicating a 99.8 per cent probability that the 22-/321-GHz
maser-line ratio varies systematically with the mass-loss rate such
that log$_{10} (L_{22}/L_{321}) = 0.32 + 0.88$\,log$_{10} (\dot
M/10^{-6}\,{\rm M}_{\odot}\,{\rm yr}^{-1})$.  This would be expected
if the maser transitions are inverted at different distances from the
central star and a similar trend was tentatively reported by Menten \&
Melnick (1991).

\subsection{The OH lines: upper limits}

1612- and 1667-MHz spectra of R~Aqr were obtained using the VLA during
1994 June. The 2-hr integration resulted in a substantially lower
noise level than that achieved by Ivison et al.\ (1994); still,
neither of the OH lines were detected. The noise level (in a region
covering $v_{\rm lsr}$ = $-96$ to $+46$\,km\,s$^{-1}$) was 2.7\,mJy
for both spectra (with 1.1-km\,s$^{-1}$ channels).

If the line tentatively identified by Ivison et al.\ (1994) at
$-14$\,km\,s$^{-1}$ were a real feature (and OH masers are usually
variable on much larger timescales than masers that form closer to the
LPV) then we would have detected it at the $\sim30$-$\sigma$
level. Our limit on the integrated flux (assuming a line width of
8.6\,km\,s$^{-1}$) is $S(3\sigma) < 0.025$\,Jy\,km\,s$^{-1}$, or $<
1.38 \times 10^{-24}$\,W\,m$^{-2}$, and we conclude that the feature
described by Ivison et al.\ (1994) was spurious.

 \begin{table*}
 \caption{Parameters for the VLA A-configuration
 observations of H1$-$36 during 1994 May 01.}
 \label{h1-36-vla}
 \begin{tabular}{ccccccc}
 \hline
Line &Velocity  &Spectral   &Beamsize and            &Central &Integrated flux               \\
     &resolution&noise (mJy)&PA ($''$ and $^{\circ}$)&velocity&density (Jy\,km\,s$^{-1}$)\\
&&&&&\\
OH -- 1612\,MHz  &0.28\,km\,s$^{-1}$&5.2&$2.2 \times 0.6, 172$ &$-124.5\pm1.0$km\,s$^{-1}$&$0.71\pm0.05$\\
OH -- 1667\,MHz  &0.27\,km\,s$^{-1}$&4.5&$2.1 \times 0.6, 172$ &$-119\pm1.0$km\,s$^{-1}$&$0.19\pm0.03$\\
H$_2$O -- 22\,GHz&1.32\,km\,s$^{-1}$&4.0&$0.29 \times 0.06, 22$&$-121\pm1.5$km\,s$^{-1}$&$0.16\pm0.04$\\
 \hline
 \end{tabular}
 \end{table*}

\section{Observations of H1$-$36}

\begin{figure}
\setlength{\unitlength}{1mm}
\begin{picture}(50,117)
\put(0,0){\includegraphics{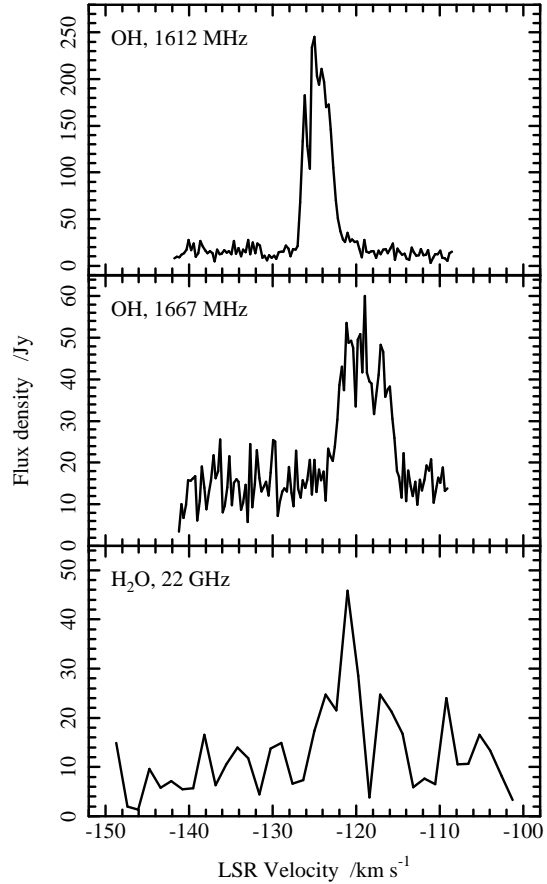}}
\end{picture}
\caption{Masers observed towards H1$-$36 using the VLA in A configuration
during 1994 May 01.}
\label{h1-36}
\end{figure}

\begin{figure}
\setlength{\unitlength}{1mm}
\begin{picture}(80,75)
\put(0,0){\includegraphics{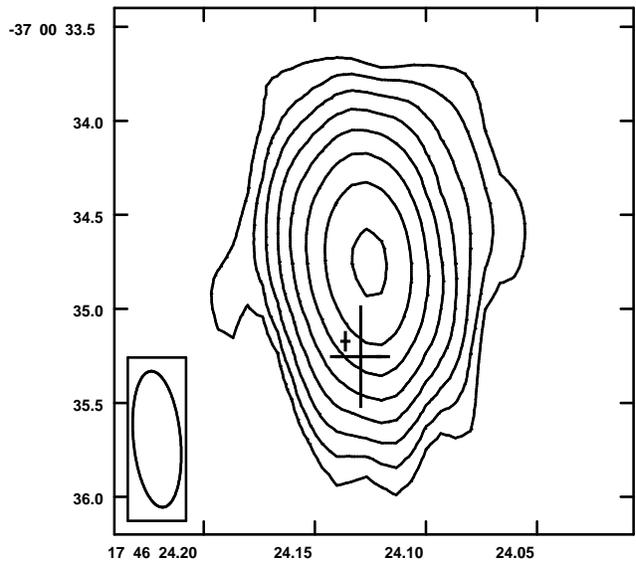}}
\end{picture}
\caption{A-configuration 8.4-GHz isophotal map of H1$-$36 obtained
during 1991 Sept). The positions of the OH and 22-GHz H$_2$O masers
are marked and the size of the restoring beam is shown. Contours are
plotted at 1, 2, 4 ... 128 times the 3-$\sigma$ level of
0.25\,mJy\,beam$^{-1}$.}
\label{xband}
\end{figure}

Fig.~\ref{h1-36} shows the OH and H$_2$O spectra of H1$-$36 obtained
during 1994 May. Both the spatial and the velocity resolution were
better than those obtained by Ivison et al.\ (1994) and SIH95 during
the survey-style observations, allowing us to rule out a thermal
origin for the OH and H$_2$O line emission ($T_{\rm b} > 9 \times
10^3$ and $3 \times 10^3$\,{\sc k}, respectively). Despite this, the
longer integration times reported here meant that the noise values in
the OH and H$_2$O spectra were several times lower than in our 1993
spectra (see Table~\ref{h1-36-vla}).

The 1612-MHz OH line centroid ($v_{\rm lsr} = -124.5 \pm
1.0$\,km\,s$^{-1}$) is displaced by $-5.5$\,km\,s$^{-1}$ from the
mainline centroid, though a faint red wing stretches to the centre of
the mainline. The full widths are $\sim8$\,km\,s$^{-1}$ for both of
the lines, whilst the FWHM are $\sim4$ and $\sim6$\,km\,s$^{-1}$ at
1612 and 1667\,MHz, respectively. The velocity of the strongest peak
at 1612\,MHz corresponds with the 1993 May and July values (to within
the uncertainties, Ivison et al.\ 1994). This is also true of the
H$_2$O line (which is at only 25 per cent of its 1993 intensity).

If the observed 1612-MHz peak is the blueward component of the usual
OH pair, then we can estimate a terminal wind velocity of
10\,km\,s$^{-1}$ from its displacement from H1$-$36's SiO and
H92$\alpha$ lines (Allen et al.\ 1989; Bastian 1992); furthermore, if
the unseen redward peak has been obscured by optically thick, ionized
gas (an idea first proposed by Allen et al.\ 1989 --- see \S5.2.3)
then $\tau \geq 3.6$ at 1612\,MHz, based on the spectral rms redward
of the observed line.

\section{On the R~Aqr system}

\subsection{Radial velocity measurements in the radio}

Several estimates of R~Aqr's orbital period have been proposed; none,
however, have been robust. Table~\ref{orb} lists the orbital
periodicities that have been suggested to date, whilst more recent
radial velocity (RV) data have been presented by Wallerstein (1986)
and Hinkle et al.\ (1989). The RVs do not strongly support any of the
suggested periodicities. Hinkle et al.\ state that the period must be
between 18 and 21\,yr or longer than 32\,yr. Most solutions for
periods of over 32\,yr require $e > 0.3$.

Monitoring the position and velocity of the R~Aqr water maser could
reveal the system's elusive binary period.  The VLA offers both high
velocity resolution and high positional accuracy with which to track a
Mira.  Table~\ref{raqr-vla} shows the measurements from our first
three epochs of monitoring; so far, there are no appreciable
positional shifts and only one appreciable velocity shift. The problem
could be that the timescale for the rise and fall of individual maser
features (which is weakly correlated with the pulsational period) is
shorter than the orbital period.  This is compounded by strong, new
maser features (such as that seen at the 0.4-Jy level in 1995 June)
which can obscure weaker features such as the 0.1-Jy peak of 1994 May
(see Fig.~\ref{vla}) .

The 8-km\,s$^{-1}$ pedestal is the most enduring feature of the maser
spectrum, and its stability at the $\Delta v_{\rm lsr} <
1$\,km\,s$^{-1}$ level over a period of two years implies a low
orbital inclination, or a high $M_{\rm giant}:M_{\rm dwarf}$ ratio, or
an orbital period that is measured in decades, which is consistent
with optical data. If the pedestal is the result of thermal emission
(at the 10-mJy level) from near the photosphere of the rotating LPV
then its FWZM translates into a lower limit for the rotational
period. The period will approach the lower limit if intrinsic line
broadening is negligable, i.e.\ if the width is due entirely to
Doppler broadening. The maximum photospheric line-of-sight velocity
implied by the pedestal is 4\,km\,s$^{-1}$, well within the scatter of
the observed rotational velocities for late-type giants ($v$\,sin\,$i
= 1 - 9$\,km\,s$^{-1}$ for K giants, De Medeiros, Melo \& Mayor
1996). This yields a rotational period of $>17$\,yr for an LPV radius
of $500\pm100$\,R$_{\odot}$ (Haniff, Scholz \& Tuthill 1995; van Belle
et al.\ 1996).

 \begin{table}
 \caption{Orbital periods suggested in the literature.}
 \label{orb}
 \begin{tabular}{ll}
 \hline
Periodicity&Description\\
/yr&\\
&\\
26.7&RV measurements (Merrill 1950). Later refuted\\
    &by Jacobson \& Wallerstein (1975).\\
24.7&Same RV data as Merrill (1950), used by\\
    &Kurochkin (1976).\\
44  &From two periods of peculiar activity in the\\
    &Mira's LC that are suggestive of its eclipse\\
    &(Willson et al.\ 1981).\\
 \hline
 \end{tabular}
 \end{table}

\subsection{Mechanisms to explain the simultaneous data}

\subsubsection{Disruption by the white dwarf secondary?}

Any periodicity in excess of 8\,yr would yield a binary separation
which would put the hot star in the H$_{2}$O maser zone (10---20\,au),
i.e., the companion star would orbit in the region normally occupied
by the masing molecular gas. The companion could then reduce
amplification by inducing turbulent motion and disrupting the
velocity-coherent paths in the outer envelope where the bright 22-GHz
masers form (Hall et al.\ 1990a, 1990b). The interaction of the
accretion flow and the Mira's wind could cause large velocity
gradients and turbulence in the circumstellar envelope, which reduce
the velocity coherence path (gain length).

\subsubsection{Consistency with SIH95 model?}

SIH95 put forward another model that appears to be consistent with the
results of not only their survey of maser-line transitions towards
symbiotic Miras, but also with the details of individual systems such
as R~Aqr and H1$-$36.  In their model, UV radiation and a fast wind
from the white dwarf companion dissociate and sweep away the dusty,
molecular gas shed by the Mira. Even the molecular material shielded
by the Mira from direct irradiation becomes dissociated as the result
of Rayleigh scattering in the Mira's wind. However, since the R~Aqr
hot component has a relatively low luminosity (and the nebula is
ionisation bounded), the effectiveness of direct and indirect
molecular dissociation by UV irradiation would be much reduced; even
on the irradiated hemisphere of the Mira, an extensive neutral cavity
may remain beyond the ionization front of the hot dwarf. Thus we have
an environment where 22-GHz H$_2$O masers may survive, in keeping with
our observations.

The energy levels of the 22-GHz and 321-GHz transitions are 643 and
1861\,K above the ground state, respectively.  Yates, Field \& Gray
(1997) suggest that 321-GHz masers are excited closer to the Mira than
22-GHz masers.  Given the absence of a suitable dusty, molecular
environment in the outer reaches of the R~Aqr Mira's wind, the
observed 22-GHz H$_{2}$O maser lines will be weak and narrow in
comparison to those observed towards field Miras.

Note also that our maps (Fig.~\ref{maps}) seem to indicate that the
22-GHz maser and the ionized gas are co-spatial (\S3.1). This makes
little sense unless the molecular gas is protected in a neutral
cavity, much closer to the LPV than would normally be anticipated. The
superimposition can be explained if the maser is within a cocoon of
ionized gas and the strongest contribution to the free-free radiation
component comes from the densest portion of the ionized wind, which is
near the ionisation front and close to the Mira.

\subsubsection{Line obscuration at low frequencies?}

This was proposed Allen et al.\ (1989), and relies on the fact that
the LPV's wind is optically thick at low frequencies.  The spectral
energy distributions of H1$-$36 and R~Aqr show the turnover to
optically thin free-free emission is at $\sim10$\,GHz (Ivison, Hughes
\& Bode 1992; Ivison et al.\ 1995b). Below 10\,GHz, we can expect the
line emission to be hidden from us by the optically thick, ionized
gas.

The maser data support Allen's idea: first, the absence of OH maser
emission towards R~Aqr (see \S3.3); second, only one component of the
normal 1612-MHz OH mainline pair is seen towards H1$-$36, which
suggests a large optical depth between the front and back of the
masing shell in the line of sight through the system.  Third, the
observed lines become progressively more intense with increasing line
frequency, suggesting lower obscuration: the OH line is absent, the
22-GHz emission is weak and the line emission at 43, 86, 299, 301 and
321\,GHz is strong. There is no evidence, however, that the opacity at
22\,GHz is sufficient to move the 22-/321-GHz line flux and luminosity
ratios closer to those typically observed towards field Miras (a
22-/321-GHz line luminosity ratio of 1.5 would require $\tau \sim 2$
at 22\,GHz).

\section{Concluding remarks}

We have presented maser-line spectra and maps of the symbiotic Miras,
R~Aqr and H1$-$36, together with contemporaneous, high-resolution
continuum maps. Our measurements have resulted in the first detection
of sub-mm water maser emission towards a symbiotic system (R~Aqr) and
we have discussed three mechanisms that could cause the observed line
properties in the geometries implied by the continuum and
spectral-line maps.

With the data available to us, it is not possible to discriminate
between three possible scenarios, although one of these requires
conditions that are not supported by other measurements. The
mechanisms that we favour inhibit the formation of OH and 22-GHz
H$_2$O masers in their natural environments, employing either the
orbital motion of the companion or its radiation field and fast wind. The
third, which reduces maser-line intensities via the free-free opacity
of circumstellar, ionized gas (and thus has an effect that is
dependent on the rest frequency of the maser transition) is somewhat
less credulous since to reduce the intensity of the 22-GHz maser line
by at least an order of magnitude relative to the 321-GHz line would
require the free-free emission to remain optically thick at that
frequency, which does not appear to be the case (see Ivison et al.\
1992, 1995b).

\bsp


\begin{thebibliography}{99} 
\bibitem{}Allen D.A., Hall P.J., Norris R.P., Troup E.R., Wark R.M., Wright A.E., 1989, MNRAS, 236, 363{}{}
\bibitem{}Bastian T.S., 1992, ApJ, 387, L77{}{}
\bibitem{}Bowers P., Hagen W., 1984, ApJ, 285, 637{}{}
\bibitem{}Chinarova L.L., Andronov I.L., Schweitzer E., 1996, in
Proc.\ IAU Symp.\ No.\ 180. Kluwer, Dordrecht, in press{}{}
\bibitem{}De Medeiros J.R., Melo C.H.F., Mayor M., 1996, A\&A, 309, 465{}{}
\bibitem{}Gray M.D., Ivison R.J., Humphreys E.M.L., Yates J.A., 1998, MNRAS, submitted{}{}
\bibitem{}Gray M.D., Ivison R.J., Yates J.A., Humphreys E.M.L., Hall P.J., Field D., 1995, MNRAS, 277, L67{}{}
\bibitem{}Hall P.J., Allen D.A., Troup E.R., Wark R.M., Wright A.E., 1990a, MNRAS, 243, 480{}{}
\bibitem{}Hall P.J., Wright A.E., Troup E.R., Wark R.M., Allen D.A., 1990b, MNRAS, 247, 549{}{}
\bibitem{}Haniff C.A., Scholz M., Tuthill P.G., 1995, MNRAS, 276, 640{}{}
\bibitem{}Hinkle K.H., Wilson T.D., Scharlach W.W.G., Fekel F.C., 1989, AJ,
98, 1820{}{}
\bibitem{}Hollis J.M., Wright M.C.H., Welch W.J., Jewell P.R., Crull Jr, H.E., Kafatos M., Michalitsianos A.G., 1990, ApJ, 361, 663{}{}
\bibitem{}Ivison R.J., Hughes D.H., Bode M.F., 1992, MNRAS, 257, 47{}{}
\bibitem{}Ivison R.J., Seaquist E.R., Hall P.J., 1994, MNRAS, 269, 218{}{}
\bibitem{}Ivison R.J., Seaquist E.R., Hall P.J., 1995a, in Watt G.D., Williams P.M., eds, Circumstellar Matter 1994, p.\ 255{}{}
\bibitem{}Ivison R.J., Seaquist E.R., Schwarz H.E., Hughes D.H., Bode M.F., 1995b, MNRAS, 273, 517{}{}
\bibitem{}Jacobson T.S., Wallerstein G., 1975, PASP, 87, 269{}{}
\bibitem{}Karovska M., Nisenson P., Belefic J., 1993, ApJ, 402, 311{}{}
\bibitem{}Kurochkin N.E., 1976, Soviet Astr.\ Lett., 2, 169{}{}
\bibitem{}Martinez A., Bujarrabal V., Alcolea J., 1988, A\&A, 74, 273{}{}
\bibitem{}Menten K.M., Melnick G.J., 1991, ApJ, 377, 647{}{}
\bibitem{}Merrill P.W., 1950, ApJ, 112, 514{}{}
\bibitem{}Schwarz H.E., Nyman L.-\AA., Seaquist E.R., Ivison, R.J., 1995, A\&A, 303, 833{}{}
\bibitem{}Seaquist E.R., Krogulec M., Taylor A.R., 1993, ApJ, 410, 260{}{}
\bibitem{}Seaquist E.R., Ivison R.J., Hall P.J., 1995, MNRAS, 276, 867 (SIH95){}{}
\bibitem{}Solf J., Ulrich H., 1985, A\&A, 148, 274{}{}
\bibitem{}Sopka R.J., Herbig G., Kafatos M., Michalitsianos A.G., 1982,
ApJ, 258, L35{}{}
\bibitem{}van Belle G.T., Dyck H.M., Benson J.A., Lacasse M.G., 1996,
AJ, 112, 2147{}{}
\bibitem{}Wallerstein G., 1986, PASP, 98, 118{}{}
\bibitem{}Wallerstein G., Greenstein J.L., 1980, PASP, 92, 275{}{}
\bibitem{}Willson L.A., Garnavich P., Mattei J.A., 1981, Inf.\ Bull.\ Var.\ Stars, No.\ 1961{}{}
\bibitem{}Yates J.A., Cohen R.J., Hills R.E., 1995, MNRAS, 273, 529{}{}
\bibitem{}Yates J.A., Field D., Gray M.D., 1997, MNRAS, 285, 303{}{}
\end{thebibliography}
\end{document}